\documentclass[12pt]{article}
\newcommand{\blind}{0}

\usepackage{times,color,verbatim,graphicx,amssymb,lscape,amsmath}
\usepackage{graphicx,graphics,psfrag,epsf}
\usepackage{enumerate}
\usepackage{url} % not crucial - just used below for the URL
\usepackage{multirow}
\usepackage{multicol}
\usepackage{rotating}
\usepackage{relsize}
\usepackage{bm}
\usepackage{hyperref}
\usepackage{multicol}
\usepackage{setspace}
\usepackage{enumerate}
\usepackage{color}
\usepackage[authoryear,sort]{natbib}
\usepackage{footnote}
\usepackage{mymacros}

\date{}
\begin{document}
\def\spacingset#1{\renewcommand{\baselinestretch}%
{#1}\small\normalsize} \spacingset{1}
%%%%%%%%%%%%%%%%%%%%%%%%%%%%%%%%%%%%%%%%%%%%%%%%%%%%%%%%%%%%%%%%%%%%%%%%%%%%%%

\if0\blind
{
  \title{\bf Penalized Weighted Least Squares for Outlier Detection and Robust Regression}
  \author{Xiaoli Gao\thanks{
    The authors gratefully acknowledge \textit{Simons Foundation (\#359337, Xiaoli Gao) and UNC Greensboro New Faculty Grant (Xiaoli Gao)}}\hspace{.2cm}\\
    Department of Mathematics and Statistics\\
     University of North Carolina at Greensboro\\
    and \\
    Yixin Fang \\
    Department of Population Health\\
    New York University School of Medicine}
  \maketitle
} \fi

\if1\blind
{
  \bigskip
  \bigskip
  \bigskip
  \begin{center}
    {\LARGE\bf Penalized Weighted Least Squares for  Outlier Detection and Robust Regression}
\end{center}
  \medskip
} \fi

\bigskip
\begin{abstract}
To conduct regression analysis for data contaminated with outliers, many approaches have been proposed for simultaneous outlier detection and robust regression, so is the approach proposed in this manuscript. This new approach is called ``penalized
weighted least squares" (PWLS). By assigning each observation an individual weight
and incorporating a lasso-type penalty on the log-transformation of the weight vector, the PWLS is able to perform
outlier detection and robust regression simultaneously. A Bayesian point-of-view of the PWLS is provided, and it is showed that the PWLS can be seen as an example of M-estimation. Two methods are developed for selecting the tuning parameter in the PWLS. The performance of the PWLS  is demonstrated via simulations and real applications.
\end{abstract}

\noindent%

\vfill

\newpage
\spacingset{1.45} % DON'T change the spacing!

\section{Introduction}

In statistics, an outlier is an observation that does not follow the model of the majority of the data.
Some outliers may be due to intrinsic variability of the data; this type of outliers should be examined carefully using some subgroup analysis. Other outliers may indicate errors such as experimental error and data entry error; this type of outliers should be down-weighted.

To conduct regression analysis for data contaminated with outliers, one can detect outliers first
and then run ordinary regression analysis using the data with the detected outliers deleted \citep{weisberg:book2005}, or run some version of robust regression analysis which is insensitive to the outliers \citep{huber:1973}. Alternatively, many approaches have been proposed to simultaneously perform outlier detection and robust regression. See for example, the least median of squares \citep{siegel:1982}, the least trimmed squares \citep{rousseeuw:1984}, S-estimates \citep{rousseeuw.yohai:1984}, Generalized S-estimates \citep{croux.rousseeuw:1994}, MM-estimates \citep{yohai:1987}, the robust and efficient weighted least squares estimators \citep{gervini.yohai:2002}, and forward search \citep{atkinson.riani.cerioli:2003}. \cite{boente.pires.rodrigues:2002} also
studied outlier detection under principal components model.
One can refer to \cite{maronna.ea:2006}  and \cite{hadi.imon.werner:2009} for broader reviews of some recent robust regression procedures and outlier detection procedures.

In this manuscript, we propose a new approach, penalized weighted least squares (PWLS). By assigning each observation an individual weight
and incorporating a lasso-type penalty on the log-transformation of the weight vector, the PWLS is able to perform
outlier detection and robust regression simultaneously. For this aim, assume the data are from the following model,
\bel{model1}
y_i=\bx_i'\bbeta^*+\veps_i/w_i^*,
\eel
where $\bx_i\in \mathbb{R}^p$ are the predictor vectors, $\bbeta^*=(\beta_1^*,\cdots,\beta_p^*)'$ is the coefficient vector, and $\veps_i$ with $E(\veps_i|\bx_i)=0$ are random errors following some unknown distribution $F(\cdot | \bx_i)$ independently. Also assume the data are contaminated with outliers, and therefore the underlying weights $w_i^*$ are introduced, with $w_i^*<1$ indicating outliers and $w_i^*=1$ indicating non-outliers. We shall start our discussion with the homogeneous setting where $Var(\veps_i)=\sigma^2$ for $1\le i\le n$, and then generalize it to the heterogeneous
setting where $\veps_i=g(\bx_i'\btheta)\eps_i$ with $E(|\eps_i|)=1$ for $1\le i\le n$.

In ordinary least squares (OLS) regression,
suspected outliers could be visualized by plotting the residual $r_i=y_i-\bx_i'\widehat{\bbeta}$
 against the predicted outcome $\widehat{y}_i$, where $\widehat{\bbeta}$ is the
 estimate of $\bbeta$, along with other visualizing tools such as studentized-residuals plot
  and Cook's distance plot \citep{weisberg:book2005}. However, when there are multiple outliers,
   these simple methods can fail, because of two phenomena, masking and swamping.
   These two phenomena can be demonstrated by examining a famous artificial dataset,
   the Hawkins-Bradu-Kass (HBK) data \citep{hawkins.ea:1984}.

\begin{figure}[tp]\label{Fig:HBK}
  \caption{An illustrative example using HBK data.
  Left: Studentized residual plot from LS regression, with threshold lines of $\pm 2.5$
  (non-outliers displayed as light dots and outliers displayed as dark asterisks);
 Middle: PLWS weight solution paths  with a selected tuning parameter shown in a
 vertical line (non-outliers displayed as light solid lines and outliers displayed as dark dashed lines);
  Right: Outlier probability paths generated using
  random weighting with a selected tuning parameter shown in a vertical line
  (non-outliers displayed as light solid lines and outliers displayed as dard dashed lines).
  %A zoom-in region for the first 10 curves
    %The grey horizontal line is   the probability for 0.5.
  }\label{fig-toy2-1}
\centering
$$
 \scalebox{0.4} [0.35]{\includegraphics{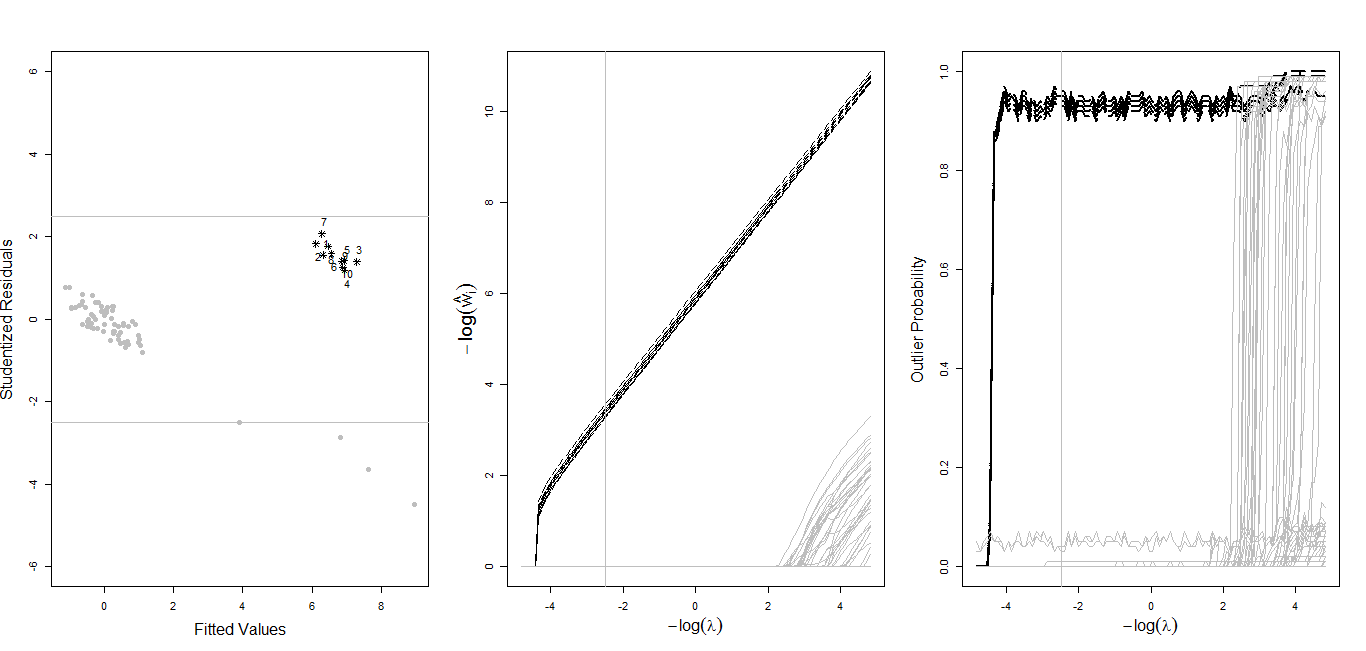}}
 $$
\end{figure}

The HBK data consist of 75 observations, where each observation has one outcome variable
 and three predictors. The first 14 observations are leverage points; however,
 only the first 10 observations are actual outliers. The studentized residual plot is
 shown in the left panel of Figure \ref{Fig:HBK}, where those from the 10 actual
 outliers are displayed differently. Those observations with large residuals bigger
 than some threshold are suspected to be outliers, and the threshold $2.5$
 suggested by \cite{rousseeuw.leroy:1987} is also shown in the residual plot.
 It is shown that three non-outliers (displayed in light dot) are beyond the
 threshold lines and therefore are suspected to be outliers; this is the swamping phenomenon.
 It is also shown that seven outliers (displayed in dark asterisk) are within the threshold
  lines and therefore survive the outlier screening; this is the masking phenomenon.

The other two plots in Figure \ref{Fig:HBK} are the outputs of our new method, the PWLS.
 As tuning parameter $\lambda$ goes from zero to infinity, it generates
 a weight path for each observation. These weight paths are shown in the middle panel;
 solution paths of non-outliers and outliers are in light solid curves and dark dashed curves,
 respectively. We can see that paths of the ten actual outliers are distant from the others,
 and at the selected tuning parameter $\widehat{\lambda}$ (displayed in a vertical line),
 weights of those non-outliers are exactly equal to one while the weights of those ten
 outliers are very small. The choice of optimal tuning parameter is
to be presented in Subsection 3.2, where a random-weighting procedure is
developed to estimate the probability
 of each observation being outlier at each value of tuning parameter $\lambda$.
  For the HBK data, such probabilities along a wide range of $\lambda$ are shown
  in the right panel; light solid curves and dark dashed curves are for non-outliers
  and outliers, respectively. We can see  that the estimated outlier probabilities
   of those ten outliers are much higher than the others, and at the same $\widehat{\lambda}$
   (vertical line), the probabilities from non-outliers are exactly equal to or at least close to $0$
   while those from ten outliers are close to $1$.

The proposal of our new method is motivated by a seminal paper, \cite{she.owen:2011}. In their paper, a regression model with a mean shift parameter is considered and then a lasso-type penalty is incorporated. However, the soft-thresholding implied by the lasso-type penalty cannot counter the masking and swamping effects and therefore they introduced a hard-thresholding version of their method. Surprisingly, this small change from soft-thresholding to hard-thresholding made their method work well for countering the masking and swamping effects, although the mysterious reason behind this was not uncovered.

The remaining manuscript is organized as follows. In Section 2, we discuss the PWLS, along with some of model justification including
its Bayesian understanding and robust investigation. In Section 3, we develop an algorithm to implement the proposed method, and two methods for selecting the tuning parameter in it. In Section 4, we extend the PWLS to heterogeneous models, in particular, the variance function linear models. In Section 5, we evaluate the performance of the newly proposed method using simulations and real applications. Some discussion is in Section 6 and the technical proof is relegated to the Appendix.

\section{The Penalized Weighted Least Squares}

If the weight vector $\bw^*=(w_1^*, \cdots, w_n^*)'$ in model \eqref{model1} is given in advance,
$\bbeta^*$ can be estimated by minimizing the weighted sum of squares, $
\sum_{i=1}^n {w_i^*}^2 (y_i-\bx_i'\bbeta)^2$. In the approach we develop here,
 we allow weights to be data-driven and estimate
 both $\bbeta^*$ and $\bw^*$ jointly by
 minimizing the following penalized weighted least squares (PWLS),
\bel{eq:pwls}
(\hbbeta, \widehat\bw)=\argmin_{\bbeta, \bw}
\left\{ \sum_{i=1}^n w_i^2 (y_i-\bx_i'\bbeta)^2 +\sum_{i=1}^n \lm |\log(w_i)|\right\},
\eel
where tuning parameter $\lambda$ controls the number of suspected outliers. The non-differentiability of penalty $|\log(w_i)|$ over $w_i=1$ implies that some of the components of $\widehat{\bw}$ may be exactly equal to one. Then the observations corresponding to
$\widehat{w}_i=1$ survive the outlier screening, while those corresponding to $\widehat{w}_i\neq 1$
are suspected to be outliers. Therefore, the PWLS can perform simultaneous outlier detection and robust estimation.

Noting that $|\log(w_i)|=|\log(1/w_i)|$, we can assume that all the components of $\bw$
are either less than one (suspected outliers) or equal to one (non-outliers). In fact, any $w_i>1$ must not be a solution since it can be always replaced by $\widetilde{w_i}=1/w_i<1$ and decreases the objective function. Therefore, in the first
term of the objective function of (\ref{eq:pwls}), those suspected outliers are assigned smaller weights than the others.

The performance of the PWLS depends crucially on the determination of tuning parameter $\lambda$, ranging from $0$ to $\infty$. When $\lambda$ is sufficiently large, all $\log(\widehat{w_i})$ become zero, and consequently all observations survive outlier screening.
When $\lambda$ is sufficiently small, some $\widehat{w}_i$ become zero, and consequently they could be suspected as outliers. Therefore, the tuning parameter selection plays an important role in determining the amount of outliers. Two methods for tuning parameter selection are discussed in the next section.

Finally, we should emphasize that the PWLS is not a variation of the classical weighted least squares (WLS; see e.g., \citealp{weisberg:book2005}) aiming for fitting heterogeneous data. The PWLS is coined because the first term in the objective function of (\ref{eq:pwls}), the weighted sum of squares, is the same as that for the WLS. We could conduct variable selection by adding some penalty term on the regression coefficients in the WLS and also call it the penalized weighted least squares, but the readers should not be confused by these names, keeping in mind that the goal of this manuscript is outlier detection rather than variable selection.

\subsection{A Bayesian understanding of the PWLS} \label{sec-bayes}

\begin{figure}[tp]
  \caption{Display of some functions. Left: Improper priors of $v_i$ with hyper-parameter
  $\lambda_0=1, 2, 3, \infty$; Middle: The $\rho$ function with tuning parameter $\lambda=1, 2, 3$.
  Right: The $\psi$ function with tuning parameter $\lambda=1, 2, 3$.
  }\label{Fig:prior-rho}
\centering
$$
  \scalebox{0.4} [0.40]{\includegraphics{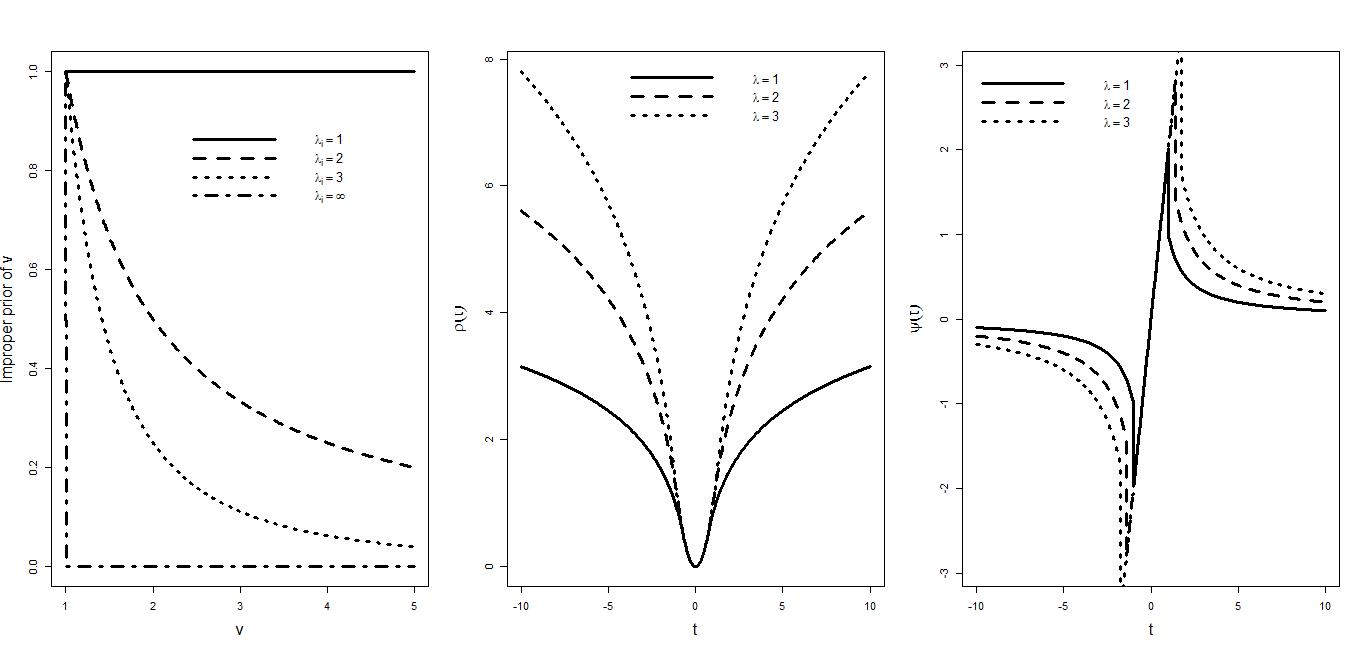}}
  $$
\end{figure}

We provide a Bayesian understanding of model \eqref{eq:pwls}. Denote $\nu_i=1/w_i$ and $\bnu=(\nu_1,\cdots, \nu_n)'$.
Let $\pi(\bbeta)$, $\pi(\sigma^2)$, and $\pi(\nu_i)$ be the independent prior distributions of
$\bbeta$, $\sigma^2$, and $\nu_i$, respectively. Assume non-informative priors $\pi(\bbeta)\propto 1$
and $\pi(\sigma^2)\propto 1/\sigma^2$ for $\bbeta$ and $\sigma^2$, respectively.
 Also assume that
$\nu_i$ has  a  Type I Pareto distribution with hyper-parameter $\lm_0 \geq 1$; that is,
\bel{eq:prior}
\pi(\nu_i)\propto{\nu_i^{1-\lm_0}} I( \nu_i\ge 1), \quad {\rm for~}1\le i\le n,
\eel
where $I(\cdot)$ is the indicator function. The prior distribution of $\nu_i$ with different hyper-parameters
 is shown in the left panel of Figure \ref{Fig:prior-rho}. In particular, it is the uniform
 non-informative prior when $\lambda_0=1$ and
  Jeffreys non-informative prior when $\lambda_0=2$.

Then the joint posterior distribution of the parameters is equal to
$$
\pi(\bbeta,\sigma^2,\bnu|\by)\propto(\sigma^2)^{-n/2-1}\prod_{i=1}^n \nu_i^{-\lm_0}
\exp\left\{-\frac{1}{2\sigma^2} \sum_{i=1}^n \frac{1}{\nu_i^2}  (y_i-\bx_i'\bbeta)^2\right\}.
$$
%whose logarithm is
%\bel{eq:log-prior}
%-(\frac{n}{2}+1)\log(\sigma^2)-\sum_{i=1}^n \lm \log( \nu_i) -\frac{1}{2\sigma^2}\left\{ \sum_{i=1}^n  \frac{1}{\nu_i^2}  (y_i-\bx_i'\bbeta)^2\right\}.
%\eel
The mode, $(\hbbeta, \widehat\bnu)$, of the above posterior distribution is
\bel{eq:mode1}
(\hbbeta, \widehat\bnu)=\displaystyle\argmin_{\bbeta, \bnu}
\left\{ \sum_{i=1}^n \frac{1}{\nu_i^2}(y_i-\bx_i'\bbeta)^2 +\sum_{i=1}^n 2\widehat\sigma^2\lm_0|\log(\nu_i)|\right\},
\eel
where $\widehat{\sigma}^2=(n+2)^{-1}\sum_{i=1}^n \widehat{\nu}_i^{-2}(y_i-\bx_i'\hbbeta)^2.$
Thus \eqref{eq:mode1} is equivalent to \eqref{eq:pwls}  if $\lm=2\widehat\sigma^2\lm_0$.

\subsection{In connection with M-estimation} \label{sec-m-est}

We demonstrate that the PWLS is an example of M-estimation by deriving its implicit $\psi$ and $\rho$ functions.
 Consider  M-estimation with $\psi$ function,
 \bel{eq-psi}
 \psi(t, \lm)=\left\{\begin{array}{ll}
  \lm/t, & {\rm if~} |t|>\sqrt{\lm/2}, \\
    2t, & {\rm if~} |t|\le\sqrt{\lm/2},
 %    -\lm/t & {\rm if~} t<-\sqrt{\lm}
      \end{array}\right.
 \eel
 and the corresponding $\rho$ function,
 \bel{eq-rho}
   \rho(t, \lm)=\left\{\begin{array}{ll}
   \lm \log(|t| \sqrt{2/\lm})+\lm/2,& {\rm if~} |t|>\sqrt{\lm/2}, \\
    t^2, & {\rm if~} |t|\le\sqrt{\lm/2}.
      \end{array}\right.
   \eel
The above $\rho$ and $\psi$ functions with different $\lambda$ are displayed in the middle panel and right panel of
Figure \ref{Fig:prior-rho}, respectively. Apparently, the proposed $\psi$ function in \eqref{eq-psi}
is non-decreasing near the origin, but decreasing toward 0 far from the origin.
Therefore, \eqref{eq-rho} generates a redescending M-estimator having special robustness properties.

%for example, as shown in the right panel of Figure \ref{Fig:prior-rho},
%$\psi(t, 1)$ looks very similar to the generalized weight function with some given parameters.

Consider the M-estimator with a concomitant scale, where the concomitant scale is added to
ensure that we can estimate $\bbeta$ and $\sigma$ simultaneously,
  \bel{eq:rho-con}
  (\hbbeta_M, \widehat\sigma_M)=\argmin_{\bbeta, \sigma}\left\{
        \sum_{i=1}^n\rho\left(\frac{y_i-\bx_i'\bbeta}{\sigma}, \lm\right)+2cn\log\sigma\right\}.
  \eel

With the proof in Appendix, we show that, for any $\lambda$ and $c$,
estimator $\hbbeta_M$ from (\ref{eq:rho-con}) is the same as that from the following PWLS with the same concomitant scale $c$,
\bel{eq:pwls-con}
(\hbbeta_P, \widehat\sigma_P, \widehat\bw_P)=\argmin_{\bbeta, \sigma, \bw}\left\{ \frac{1}{\sigma^2}
\left[\sum_{i=1}^n w_i^2 (y_i-\bx_i'\bbeta)^2 +\sum_{i=1}^n\lm |\log(w_i)| \right]  +2cn\log\sigma\right\}.
\eel

  \begin{theorem}\label{thm:m-est}
  For any given $\lambda$ and $c$, the M-estimator $\hbbeta_M$  from \eqref{eq:rho-con} is the same as the
  PWLS estimator $\hbbeta_P$ from \eqref{eq:pwls-con}.
  \end{theorem}

\subsection{The adaptive PWLS}

Since the lasso was proposed by \cite{tibshirani:1996}, many other sparsity inducing penalties have also been proposed. Among them, the adaptive lasso proposed by \cite{zou:2006} has become very popular recently, partly because of its convexity, selection consistency and oracle property. If we see the penalty in \eqref{eq:pwls} as a lasso-type penalty, then we propose the following adaptive version of the PWLS (aPWLS),
\bel{eq:apwls}
(\hbbeta, \widehat\bw)=\argmin_{\bbeta, \bw}
\left\{ \sum_{i=1}^n w_i^2 (y_i-\bx_i'\bbeta)^2 +\sum_{i=1}^n \lm \varpi_i|\log(w_i)|\right\},
\eel
where $\lambda$ is a tuning parameter controlling the number of suspected outliers
and  $\bvarpi=(\varpi_1,\cdots, \varpi_n)'$ includes pre-defined penalty scale factors for all observations.
In particular, we expect $\varpi_i$ to be larger for potential outliers and smaller for normal observations.
For example, we can take $\varpi_i=1/|\log(w^{(0)}_i)|$, where $w_i^{(0)}$ are some initial estimates of $w_i$.

The selection of initial estimates $w_i^{(0)}$ is important and therefore we try to make the selection process as objective as possible. First, we obtain $w_i^{(0)}$ using \eqref{eq:pwls} with $\lambda^{(0)}=2(\widehat{\sigma}^{(0)})^2$. This tuning parameter is suggested in Subsection \ref{sec-bayes} assuming the uniform non-informative prior of \eqref{eq:prior}. Then we propose to consider $\varpi_i=1/|\log(w_i^{(0)})|$, with the convention that $1/0$ equals some large number, say $999$. Based on our limited numerical experience, we find that the performance of the PWLS is robust to a wide range of $\lm^{(0)}$, as long as the proportion of $w_i^{(0)}=1$ in the resulting $\bw^{(0)}$ is not very high (i.e., as long as it is smaller than 1 minus the proportion of ``suspected" outliers).

\section{Implementation and Tuning}

\subsection{Algorithms} \label{sec-algorithm}

%In this section, we provide algorithms for \eqref{eq:pwls} and
%\eqref{eq:pwls-appr}.
We describe an algorithm to implement the aPWLS, of which the PWLS is a specification with $\varpi_i=1$. Note that the objective function in \eqref{eq:apwls} is  bi-convex; For a given $\bw$, the function of $\bbeta$ is a convex optimization problem, and the vice versa. This biconvexity guarantees that the algorithm has promising convergence properties \citep{gorski-pfeuffer.klamroth:2007}. The algorithm is summarized in the following flow-chart.

\bigskip

\begin{tabular}{l}
  \hline
{\bf Algorithm 1} The PWLS\\
  \hline
{\bf Given} $\bX\in \mathbb{R}^{n\times p}$, $\by\in \mathbb{R}^n$, initial estimates $\bbeta^{(0)}$, $\bw^{(0)}$ and
           penalty scales $\varpi$.\\
      \quad   For any given $\lm$ in a fine grid, let   $j=1$.\\
{\bf While} not converged {\bf do}\\
 \quad  $[{\rm update}$ $\bbeta$] \\
  \quad  \quad   $\by^{{\rm adj}}=\bw^{(j-1)}\cdot\by$, $\bX^{{\rm adj}}=\bw^{(j-1)}\cdot\bX$,
              $\bbeta^{(j)}=(\bX^{{\rm adj}'}\bX^{{\rm adj}})^{-1}\bX^{{\rm adj}'}\by^{{\rm adj}}$\\
\quad     $[{\rm update}$ $\bw$] \\
 \quad   \quad      $\br^{(j)}=\by-\bX\bbeta^{(j)}$,\\
\quad   \quad     If $|r_i^{(j)}|>\sqrt{0.5\lm_i}$, let $\bw_i^{(j)}\leftarrow \sqrt{0.5\lm_i}/|r_i^{(j-1)}|$\\
 %(or $0.5\lm_i/(r_i^{(j-1)})^2$ for \eqref{eq:pwls-appr})\\
% or\\
%             \quad    \quad \quad \quad \quad\quad \quad \quad \quad let  for \eqref{eq:pwls} \\
           \quad      \quad \quad  otherwise $\bw_i^{(j)}\leftarrow 1$\\
 \quad   \quad     converged $\leftarrow \|\bw^{(j)}-\bw^{(j-1)}\|_\infty<\eps$\\
\quad  $j\leftarrow j+1$\\
 {\bf end while}\\
 deliver $\hbbeta=\bbeta^{(j)}$ and
     $\widehat \bw=\bw^{(j)}$. \\
  \hline
\end{tabular}

\bigskip

In addition, the corresponding R codes are available at \url{https://sites.google.com/a/uncg.edu/xiaoli-gao/home/r-code}. The algorithm for \eqref{eq:apwls} is illustrated in the middle panel of Figure \ref{Fig:HBK} using the HBK data, where the paths of $\widehat{\bw}$ as $\lambda$ changes are displayed.

\subsection{Tuning parameter selection} \label{sec-tuning}

We propose two methods for selecting the tuning parameter in the aPWLS; one is Bayesian Information Criterion (BIC; \cite{schwarz:1978}) and the other is based on stability selection \citep{sun.ea:2013}. Both methods are used to select an ``optimal"
  $\widehat{\lambda}$ from a fine grid of $\lambda$.

  Let $\widehat{\bbeta}(\lambda)$ and
   $\widehat{\bw}(\lambda)$ be the resulting estimates for given $\lambda$.
The BIC method chooses the optimal $\widehat{\lm}$ that minimizes
\bel{eq-bic}
{\rm BIC}(\lm)=(n-p)\log\left\{\|\widehat\bw(\lm)\cdot(\by-\bX\hbbeta(\lm))\|_2^2/\|\widehat\bw(\lm)\|_2^2\right\}+\widehat k(\lm)\{\log(n-p)+1\},
\eel
where ``$\cdot$'' is a dotted product and
 $\widehat k(\lm)=\#\{1\le i\le n: \widehat w_i(\lm)<1\}$. The first term in \eqref{eq-bic} measures the goodness of fit, and the second term measures the model complexity, where $\widehat k(\lm)$ is the number of ``outliers" detected using the current tuning parameter $\lm$. The BIC formula
indicates a trade-off between the goodness of fit and the number of suspected outliers, with smaller $\lm$ leading to more suspected outliers and vice versa. A very similar formula of BIC was also used by \cite{she.owen:2011} for the tuning parameter selection in their methods.

The stability selection method is motivated by a notion that an appropriate
tuning parameter should lead to stable outputs if the data are perturbed.
In the aPWLS, one of the main outputs is which of $n$ observations
 are suspected outliers. That is, given $\lambda$, inputting data $\mathcal{Z}$
 outputs a subset, $\mathcal{O}(\lambda; \mathcal{Z})$, consisting of all the
 suspected outliers. If there are two perturbed datasets, $\mathcal{Z}^{*1}$
 and $\mathcal{Z}^{*2}$, we hope that the two outputs,
 $\mathcal{O}(\lambda; \mathcal{Z}^{*1})$ and $\mathcal{O}(\lambda; \mathcal{Z}^{*2})$,
  be similar. Otherwise, if $\mathcal{O}(\lambda; \mathcal{Z}^{*1})$ and
   $\mathcal{O}(\lambda; \mathcal{Z}^{*2})$ are very different, then neither
    of the two outputs is trustful. Therefore, we attempt to select a $\lambda$
    such that the resulting output $\mathcal{O}(\lambda; \mathcal{Z})$ is most
    stable when $\mathcal{Z}$ is perturbed.

Before describing the stability selection method, we should first decide which
perturbation procedure is appropriate for our setting. There are three popular
perturbation procedures \citep{shao.tu:1995}: data-splitting, bootstrap, and
random weighing. Both data-splitting and bootstrap have been widely used for
constructing stability selection methods. For example, \cite{meinshausen.buhlmann:2010}
and \cite{sun.ea:2013} used data-splitting for their proposals of stability selection,
while \cite{bach:2004} used bootstrap for his proposal of stability selection. However,
neither data-splitting nor bootstrap is suitable for our purpose of outlier detection,
because any perturbed dataset using either data-splitting or bootstrap leaves out some
observations, whose statuses of being suspected outliers are unobtainable. Therefore,
we propose to use random weighting as the perturbation procedure in the construction of our stability selection method. Here the random weighting method, which is a resampling method acting like the bootstrap, is not a Bayesian method, although it was called the Bayesian bootstrap in \cite{rubin:1981}.

Let $\omega_1, \cdots \omega_n$ be some i.i.d.~random weights with $E(\omega_i)=Var(\omega_i)=1$,
 and $\bomega=(\omega_1, \cdots \omega_n)'$. Those moment conditions
  on the random weights are standard \citep{fang.zhao:2006}. With these random weights,
  we obtain the corresponding perturbed estimates,
\bel{eq:pwls-rw}
(\hbbeta(\lambda; \bomega), \widehat\bw(\lambda; \bomega))=\argmin_{\bbeta, \bw}
\left\{ \sum_{i=1}^n \omega_i w_i^2 (y_i-\bx_i'\bbeta)^2 +\sum_{i=1}^n\lm \varpi_i|\log(w_i)|\right\}.
\eel

Via (\ref{eq:pwls-rw}), any two sets of random weights, $\bomega_1$ and $\bomega_2$, give two perturbed weight estimates
 $\widehat\bw(\lambda; \bomega_1)$ and $\widehat\bw(\lambda; \bomega_2)$, which claim two sets of suspected outliers,
  $\mathcal{O}(\lambda; \bomega_1)$ and $\mathcal{O}(\lambda; \bomega_2)$. The agreement of these two sets of suspected outliers can be measured by Cohen's kappa coefficient \citep{cohen:1960}, $\kappa(\mathcal{O}(\lambda; \bomega_1), \mathcal{O}(\lambda; \bomega_2))$.
%  More details can be found in Section \ref{sec-tuning}.
  %The definition of kappa coefficient and other details of stability selection are relegated to the Appendix.

Finally, if we repeatedly generate $B$ pairs of random weights, $\bomega_{b1}$ and $\bomega_{b2}$, $b=1, \cdots, B$, we can estimate the stability of the outlier detection by
\begin{equation}
\widehat{S}(\lm)=\frac{1}{B}\sum_{b=1}^B \kappa\left(\mathcal{O}(\lambda; \bomega_{b1}), \mathcal{O}(\lambda; \bomega_{b2})\right),\label{stab}
\end{equation}
and then select $\widehat{\lambda}$ that maximizes $\widehat{S}(\lm)$. As a byproduct and without extra computing, the proposed stability selection method can provide, for each observation, an estimate for the probability of it being an outlier as $\lambda$ changes,
\begin{equation}
\widehat{P}^o_i(\lambda)=\frac{1}{2B}\sum_{b=1}^B \sum_{k=1}^2 I\left\{i\in \mathcal{O}(\lambda; \bomega_{bk})\right\}.\label{prob-outlier}
\end{equation}

The stability selection method is illustrated in Figure \ref{Fig:HBK} using the HBK data. The vertical lines shown in the middle and right panels of Figure \ref{Fig:HBK} are corresponding to $\widehat{\lm}$ selected by the stability selection method. The right panel of Figure \ref{Fig:HBK} shows the outlier probability curves using (\ref{prob-outlier}), where the curves of those ten outliers can be distinguished easily from the others.

\section{Extension of the PWLS to Heterogeneous Models}\label{sec-hetero}

Hitherto we consider  $\veps_i$ in model
  (\ref{model1}) to be homogeneous and propose the PWLS approach for simultaneously
 conducting robust regression and detecting outliers.
 However, when $\veps_i$  are also heterogeneous, and if the heterogeneity is not taken into account, some non-outliers
  with large underlying variances might be suspected falsely as outliers (the swamping phenomenon),
  while some outliers with small underlying variances might survive the outlier screening (the masking phenomenon).
  Therefore, we extend our proposal to be applicable to heterogeneous models.

Consider a heterogeneous case where $\veps_i=g(\bx_i'\bvartheta)\eps_i$ with
$E(\eps_i)=0$ and $E(|\eps_i|)=1$ for $1\le i\le n$. Here we assume $g(v)$ is a known function;
for example, $g(v)=|v|$ or $g(v)={\rm exp}(v)$. This is a broad class of heterogeneous models
considered in a seminal paper, \cite{davidian.carroll:1987}, where they proposed a general
framework for estimating parameter $\bvartheta$ in variance function $g(\bx_i'\bvartheta)$.
We refer to this class of models as variance function linear models (VFLMs).

Motivated by \cite{davidian.carroll:1987}, many authors have attempted to broaden the class of VFLMs;
to name just a few, \cite{carroll.ruppert:1988}, \cite{hall.carroll:1989}, \cite{carroll.hardle:1989},
\cite{carroll:2003BIOC}, \cite{ma.chiou.ea:2006}, and \cite{ma.zhu:2012}. Most recently, \cite{Lian.Liang.Carroll:2013}
 studied the variance function partially linear single index models (VFPLSIMs),
 in which variance function is a function of the sum of linear and single index functions;
 that is $g(\bx_i'\bvartheta+h(\bx_i'\bzeta))$, where $g$ is known and $h$ is unknown.
 In this manuscript, we demonstrate that we can extend the aPWLS to the VFLMs.
 Similarly, we can also extend the PWLS to broader and more flexible classes, say the VFPLSIMs,
 by replacing $g(\bx_i'\bvartheta)$ in the following discussion by $g(\bx_i'\bvartheta+h(\bx_i'\bzeta))$.

Without considering outliers, one of the several approaches proposed in \cite{davidian.carroll:1987}
to estimating $\bbeta$ and $\bvartheta$ in the VFLMs is described briefly in three steps.
(1) Obtain initial estimate $\widehat{\bbeta}^{\rm{homo}}$ for $\bbeta$ ignoring heterogeneity; (2) Let $R_i=|y_i-\bx_i'\widehat{\bbeta}^{\rm{homo}}|$ be the absolute residuals and obtain an estimate for $\bvartheta$, $\hbvartheta=\displaystyle\argmin_{\bvartheta} \displaystyle\sum_{i=1}^n  \left( R_i-g(\bx_i'\bvartheta)\right)^2$; (3) Obtain an updated estimate for $\bbeta$,  $\hbbeta=\argmin_{\bbeta} \sum_{i=1}^n ( y_i-\bx_i'\bbeta)^2/g^2(\bx_i'\hbvartheta)$.
  \cite{davidian.carroll:1987} showed the consistency and efficiency of this method
  under some regular conditions. Following this approach, we extend the PWLS to
  the VFLMs for robust regression and outlier detection.

With considering outliers in fitting the VFLM, the extended aPWLS has also three steps.
First, ignoring heterogeneity, obtain an initial estimate of $\hbbeta$, $\widehat{\bbeta}^{\rm{homo}}$
 using the original aPWLS proposed in Section 2;
Second, letting $R_i=|y_i-\bx_i'\widehat{\bbeta}^{\rm{homo}}|$ be the absolute residuals, obtain an estimates for $\bvartheta$,
\begin{equation}\label{eq:theta}
\hbvartheta=\argmin_{\bvartheta} \sum_{i=1}^n  \left( R_i-g(\bx_i'\bvartheta)\right)^2 ;
\end{equation}
Finally, obtain an updated estimate for $\bbeta$ and an estimate for $\bw$ via,
\begin{equation}\label{eq:update-beta}
(\hbbeta,\widehat\bw)=\argmin_{\bbeta, \bw}\left\{ \sum_{i=1}^n w_i^2 (y_i-\bx_i'\bbeta)^2/g^2(\bx_i'\hbvartheta)
+\sum_{i=1}^n\lm \varpi_i|\log(w_i)|\right\}.
\end{equation}

\begin{figure}[tp]
  \caption{An illustrative example of applying the extended aPWLS to a heterogeneous dataset. Left: Scatter-plot of $y_i$ against $g(\bx_i'\btheta)$ (non-outliers displayed as light circle dots and outliers as dark asterisks). Middle: Plot of Studentized residuals, with threshold lines of $\pm 2.5$ . Right: Weight paths from the extended aPWLS method (non-outliers displayed as light solid lines and outliers as dark dashed lines), with a selected tuning parameter shown in a vertical line.}\label{Fig:hetero}
\centering
$$
  \scalebox{0.4} [0.35]{\includegraphics{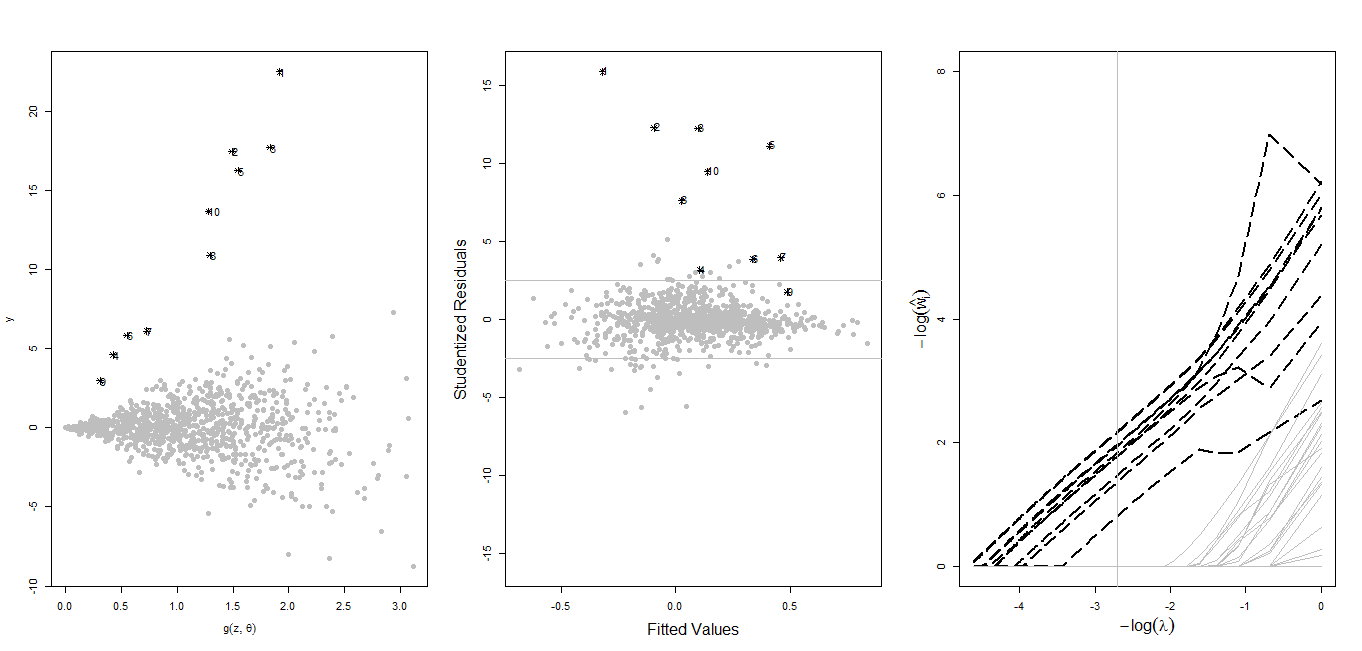}}
  $$
\end{figure}

We illustrate this extended aPWLS using a heterogeneous dataset generated from Example \ref{example:hetero1} described in the next section, where there are 1000 observations
and among them 10 observations are outliers. The scatter-plot of $y_i$ vs. $g(\bx_i'\bvartheta)$
is shown in the left panel of Figure \ref{Fig:hetero}, where 10 outliers are displayed in dark asterisks.
The studentized residuals are shown in the middle panel of Figure \ref{Fig:hetero}.
Using threshold 2.5, one outlier is not detected (the masking phenomenon) and many non-outliers
 are detected falsely as outliers (the swamping phenomenon). The weight paths from the extended
  aPWLS are shown in the right panel of Figure \ref{Fig:hetero}, where the weight paths of 10 outliers (displayed in dark dashed lines) are distinguished from the other from non-outliers (displayed in light solid lines). Moreover, at the selected tuning parameter (displayed in a vertical line), the weights from non-outliers are exactly equal to one while those from outliers are near zero.

\section{Numerical Results}

As discussed in Section \ref{sec-tuning}, tuning parameter selection
plays an important role in the penalization approach.
We first conduct some simulation studies to compare the two tuning methods, BIC and random weighting (RW).
The results show that they perform similarly in terms of outlier detection; the results are omitted here. Such a phenomenon
is also observed in \cite{sun.ea:2013}. Therefore, because BIC was used with HIPOD in \cite{she.owen:2011}, which is the main method with which our method is compared, we use BIC in all the simulation studies presented here.
However, RW is used in all three real data applications, because of the byproduct of using the random weighting method, that is, for each observation, we can visualize the probability of it being an outlier as tuning parameter $\lambda$ changes.

\subsection{Simulation studies} \label{sec-simulation}

We conduct simulation studies to demonstrate the performance of the PWLS for
outlier detection under two scenarios, homogeneous models and heterogeneous models. The PWLS is compared with the hard-IPOD (HIPOD) of \cite{she.owen:2011}.
 %and the FS of \cite{atkinson.riani.cerioli:2010}.
 In \cite{she.owen:2011}, the HIPOD was compared with four other robust regression methods. Because Example 1 we consider here is adopted from \cite{she.owen:2011}, we are able to compare the PWLS with the HIPOD and at the same time with those four robust regression methods, by combining the results presented here and those presented in \cite{she.owen:2011}.

 \begin{example}\label{example:homo1} {\rm(Homogeneous model)}
Data are generated from the mean shift model,
\bel{sim-model1}
 \by_i=\bx_i'\bbeta+\gamma_i+\veps_i,\quad 1\le i\le n,
 \eel
where $\veps_i\sim N(0,1)$ are independently and
 $\bbeta=\bone_p=(1, \cdots, 1)'$.  The first $k$ observations are set to be outliers by
letting $\gamma_i=r$ for $1\le i\le k$ and $0$ otherwise.
A matrix is generated from $\bX=(\bx_1, \cdots, \bx_n)'=\bU\bSigma^{1/2}$,
where $\bU=(u_{ij})_{n\times p}$ with $u_{ij}\sim \rm{Unif}(-15, 15)$ and $\bSigma_{ij}=(\sigma_{ij})_{p\times p}$ with
$\sigma_{ii}=1$ and $\sigma_{ij}=0.5$  for $i\neq j$.
 The design matrix is either $\bX$ (no $L$) or $\bX$ with its first $k$ rows
replaced by $L\cdot \bone_{p}$ for some positive $L$. Thus, for the former case (no $L$),
the first $k$ observations are outliers but not leverage points,
whereas for the latter case, the first $k$ observations are both leverage points and outliers.
Set  $p\in \{15, 50\}$,  $k\in \{100, 150, 200\}$,  $r=5$, and $L\in \{25, 15, 10\}$.
%Two settings are considered: (1) $n=1000$ with 100 outliers; (2) $n=500$ with 75 outliers.
\end{example}

%This data generating model is adopted from the simulation model considered in \cite{she.owen:2011}.
 The tuning parameters in the HIPOD and PWLS are selected via BIC over a grid of 100 values of $\lambda$
changing from $\lambda_{\max}$ for which no outlier is detected
 down to $\lambda_{\min}$ at which at least 50\% of the observation to be selected as outliers.
 Here we set
$\lambda_{\max}=\|(\bI_n-\bH)\by/\sqrt{\diag(\bI_n-\bH)}\|_{\infty}$,
where $\bH$ is the hat matrix and the division is elementwise, and
compute solutions along a grid 100 $\lm$  values that are equally spaced on the log scale.
The initial estimate of $\bbeta^{(0)}$ is obtained from
 \emph{lmRob()} in the R package  \emph{robust}.
The computation of the initial estimate of $w_i^{(0)}$ are described in Section \ref{sec-algorithm}.
In particular,
letting $\lm_0=\|\by-\bX\bbeta^{(0)}\|^2/(n-p)$, then
$w_i^{(0)}=\lm_0/r_i^2$ or $1$ if $\lm_0<r_i^2$, or otherwise.
%In addition, we also try to compare the PWLS with the FS in case 2 and 3,
%(obtained from the \emph{fwdlm} function in  the R package \emph{forward}).
%We could not report the FS result in Case 1 since
%the \emph{fwdlm} function fails for large sample size.
The simulation results are summarized from 1,000 iterations and they are reported in Table \ref{table-outlier-homo}.

Similar to \cite{she.owen:2011}, we evaluate the outlier detection performance using
the mean masking probability (M: fraction of undetected true outliers),
the mean swamping probability (S: fraction of non-outliers labeled as outliers),
and  the joint outlier detection rate (JD: fraction of repetitions with 0 masking) to summarize the results from the 1,000 repetitions.
The higher JD is, the better; the smaller M and S are, the better.

From Table \ref{table-outlier-homo}, we see that the PWLS outperforms the
HIPOD in all settings in terms of criteria M and JD; the PWLS has much higher
joint outlier detection rate and smaller masking probability. However,
the PWLS has a little bit bigger swamping probability measured by S.
The comparison is striking when the leverage effect is
large ($L=25)$ under 10\% outlier ratio ($k=100$).
% the HIPOD does not perform well, whereas the PWLS can still detect more than 50\% of outliers jointly.

Both PWLS and HIPOD loses their efficiency with the existence of
 a large amount of large leverage points exist ($k=200$ and $L=25$).
 However, PWLS still performs better than the HIPOD especially when the leverage value is is not too big such as ($k=200$ and $L=10$).

\begin{table}[htp]
\begin{center}
 \caption{Example \ref{example:homo1} $-$ Outlier detection evaluation for homogeneous model
 (M: the mean masking probability; S: the mean swamping probability;
  JD: the joint outlier detection rate)  }\label{table-outlier-homo}
\begin{tabular}{lllccccccc}\\ \hline\hline
 %   &&\multicolumn{8}{c} {Homogeneous Model}\\
  $k$ & $p$& Method  & JD  (\%) & M  (\%) & S  (\%) && JD (\%) & M (\%) & S (\%) \\ \hline
 % \multirow{12}{*}{\rotatebox[origin=c]{90}{Case 1}}
  \multirow{12}{*}{\rotatebox[origin=c]{0}{$100$}}
    &&& \multicolumn{3}{c} {$L=25$}&&\multicolumn{3}{c}   {$L=15$}\\
                \cline{4-6}        \cline{8-10}
 & \multirow{6}{*}{$15$}
     & PWLS   &62 &3.5  & 3.0  &&70 & 0.4 &2.9   \\
   & & HIPOD & 14& 82.1  & 0.5 &&47 & 0.9 &1.1 \\
 %  & & FS & -& -  &-&&- & - &- \\
    \cline{4-10}
    &&& \multicolumn{3}{c} {$L=10$} &&\multicolumn{3}{c}   {No $L$}\\
                \cline{4-6}        \cline{8-10}
      & & PWLS&71 & 0.4 &2.7    & &70 &0.4  & 2.6 \\
      &  & HIPOD&47 & 0.8 &1.1    & &45 &  0.9  &1.1\\
     %      & & FS & -& -  &-&&- & - &- \\
        \cline{3-10}
    &&& \multicolumn{3}{c} {$L=25$}&&\multicolumn{3}{c}   {$L=15$}\\
                \cline{4-6}        \cline{8-10}
 & \multirow{6}{*}{$50$}
     & PWLS   &58 &7.1  & 2.6  &&73 & 0.4 &2.7   \\
    & & HIPOD & 4& 88.8  & 0.5 &&49 & 0.8 &1.2 \\
    %     & & FS & -& -  &-&&- & - &- \\
        \cline{4-10}
    &&& \multicolumn{3}{c} {$L=10$}&&\multicolumn{3}{c}   {No $L$}\\
                \cline{4-6}       \cline{8-10}
    &   & PWLS&72 & 0.4 &2.4    & &57 &0.6  & 2.3 \\
     &   & HIPOD&46 & 0.8 &1.2    & &36 &  1.0  &1.2\\
    %      & & FS & -& -  &-&&- & - &- \\ \hline

\cline{2-10}
 \multirow{12}{*}{\rotatebox[origin=c]{0}{ $150$}}
    &&& \multicolumn{3}{c} {$L=25$}&&\multicolumn{3}{c}   {$L=15$}\\
                \cline{4-6}        \cline{8-10}
 & \multirow{6}{*}{$15$}
     & PWLS   &30 &53.4  & 5.5  &&68 & 0.3 &5.1   \\
   & & HIPOD & 0& 99.5  & 0.5 &&30 & 18.6 &1.1 \\
 %  & & FS & -& -  &-&&- & - &- \\

    \cline{4-10}
    &&& \multicolumn{3}{c} {$L=10$} &&\multicolumn{3}{c}   {No $L$}\\
                \cline{4-6}        \cline{8-10}
      & & PWLS&72 & 0.2 &5.1    & &68 &0.3  & 5.1 \\
      &  & HIPOD&38 & 0.9 &1.3    & &34 &  0.9  &1.4\\
     %      & & FS & -& -  &-&&- & - &- \\
        \cline{3-10}
    &&& \multicolumn{3}{c} {$L=25$}&&\multicolumn{3}{c}   {$L=15$}\\
                \cline{4-6}        \cline{8-10}
 & \multirow{6}{*}{$50$}
     & PWLS   &28 &61.3  & 5.6  &&76 & 0.2 &5.7   \\
    & & HIPOD & 0& 99.4  & 0.6 &&32 & 39.8 &1.1 \\
    %     & & FS & -& -  &-&&- & - &- \\
        \cline{4-10}
    &&& \multicolumn{3}{c} {$L=10$}&&\multicolumn{3}{c}   {No $L$}\\
                \cline{4-6}       \cline{8-10}
    &   & PWLS&84 & 0.1 &5.9    & &66 &0.2  &5.7 \\
     &   & HIPOD&50 & 0.6 &1.6    & &28 & 0.9  &1.5\\
    %      & & FS & -& -  &-&&- & - &- \\ \hline
        \cline{3-10}

  \cline{2-10}
 \multirow{12}{*}{\rotatebox[origin=c]{0}{ $200$}}
    &&& \multicolumn{3}{c} {$L=25$}&&\multicolumn{3}{c}   {$L=15$}\\
                \cline{4-6}        \cline{8-10}
 & \multirow{6}{*}{$15$}
     & PWLS   &0 &53.4  & 5.5  &&30 & 49.7 &7.6   \\
   & & HIPOD & 0& 99.5  & 0.5 &&0 & 99.8 &0.2 \\
 %  & & FS & -& -  &-&&- & - &- \\

    \cline{4-10}
    &&& \multicolumn{3}{c} {$L=10$} &&\multicolumn{3}{c}   {No $L$}\\
                \cline{4-6}        \cline{8-10}
      & & PWLS&56 & 0.5 &3.3    & &58 &0.4  & 3.1 \\
      &  & HIPOD&40 & 8.6 &1.7    & &46 &  0.7  &1.9\\
     %      & & FS & -& -  &-&&- & - &- \\
        \cline{3-10}
    &&& \multicolumn{3}{c} {$L=25$}&&\multicolumn{3}{c}   {$L=15$}\\
                \cline{4-6}        \cline{8-10}
 & \multirow{6}{*}{$50$}
     & PWLS   &0 &93.5  & 7.6  &&16 & 52.8 &6.9   \\
    & & HIPOD & 0& 99.9  & 0.7 &&0 & 99.9 &0.3 \\
    %     & & FS & -& -  &-&&- & - &- \\
        \cline{4-10}
    &&& \multicolumn{3}{c} {$L=10$}&&\multicolumn{3}{c}   {No $L$}\\
                \cline{4-6}       \cline{8-10}
    &   & PWLS&46 & 0.8 &2.2    & &24 &0.7  &2.2 \\
     &   & HIPOD&46 & 20.1 &1.9    & &24 & 0.7  &2.3\\
    %      & & FS & -& -  &-&&- & - &- \\ \hline
        \cline{3-10}

             \hline\hline

\end{tabular}
\end{center}
\end{table}

\begin{example} \label{example:hetero1} {\rm(Heterogeneous model)}
Data are generated from the following VFLM,
\bel{sim-mod2}
\by_i=\bx_i'\bbeta+\gamma_i g(\bz_i'\bvartheta)+\veps_i, 1\le i\le n,
\eel
where  $\bbeta=\bone_p$,  $\bz_i=(1, x_{ip})'$,  $\bvartheta=(1, 0.7)'$
 $\veps_i=g(\bz_i'\bvartheta)\eps_i$ with $\eps_i\sim N(0, 0.5\pi))$.
 %Note that $E(|\veps_i|)= |\bz_i'\btheta|$.
 All $\bx_i$ are generated independently from a multivariate
normal distribution $N(\bzero_p, \bSigma)$, where
$\bSigma=(\sigma_{ij})_{p\times p}$ with $\sigma_{ij}=0.5^{|i-j|}$.
 The first $k$ observations are set bo be outliers
 by letting $\gamma_i=r$ for $1\leq i\leq k$ and $0$ otherwise,
Set $n=1000$, $p\in\{15, 50\}$, $k=10$, and $r=\{20, 15, 10, 5\}$.
We consider the following two  heterogeneous settings.
\begin{itemize}
\item[] Case 1: Set $g(v)= |v|$, and this true variance function is applied in the H-PWLS.
\item[] Case 2: Set $g(v)=e^{|v|}$, but a mis-specified variance function
             $g(v)=\sqrt{|v|}$ is used.
\end{itemize}
\end{example}

We evaluate the performance of the extended PWLS for heterogeneous model (H-PWLS)
by comparing it with the PWLS and the HIPOD. For the H-PWLS, we use $\hbvartheta^{(0)}=(1,0)$ as an initial estimate
of $\btheta$ for solving (\ref{eq:theta}), and we use the same way to determine initial values $\bbeta^{(0)}$
and $\bw^{(0)}$ and select tuning parameter $\widehat{\lambda}$ for solving (\ref{eq:update-beta}) as in the PWLS.
We repeat the simulation 1,000 times and use the same measurements, JD, M, and S, to summarize the
results from these repetitions. These results are reported in Table \ref{table-outlier-hetero}.

\begin{table}[htp]
\begin{center}
 \caption{Example \ref{example:hetero1} $-$ Outlier detection evaluation for heterogeneous model
 (M: the mean masking probability; S: the mean swamping probability;
  JD: the joint outlier detection rate)}\label{table-outlier-hetero}
\begin{tabular}{lllccccccc}
\\ \hline\hline
   % &&\multicolumn{10}{c} {Heterogeneous Model}\\
 & $p$& Method  & JD (\%) & M (\%) & S (\%)  && JD (\%) & M (\%) & S (\%)\\ \hline
   \multirow{16}{*}{\rotatebox[origin=c]{90}{Case 1}}
       & && \multicolumn{3}{c} {$r=20$}&&\multicolumn{3}{c}   {$r=15$}\\
                \cline{4-6}        \cline{7-10}
 & \multirow{8}{*}{$15$}
     &   H-PWLS   & 94&0.7 &0.6   &&91 &1.0 & 0.7 \\
    & & PWLS  &35 &10.4 & 1.6       &&26 & 13.3& 1.8 \\
    & & HIPOD & 39&9.1 & 2.8   &&36 &9.7 & 4.9 \\  \cline{4-10}
   &  && \multicolumn{3}{c} {$r=10$}&&\multicolumn{3}{c}   {$r=5$}\\
                \cline{4-6}        \cline{7-10}
         & & H-PWLS & 85&1.7 & 0.8  &&20 &15.2 &1.5  \\
      &  & PWLS & 13&18.9 & 2.3       && 2 &34.8 & 4.6 \\
      &  & HIPOD & 24&13.9 & 6.0       &&3 &31.9 & 5.9 \\     \cline{3-10}
   &&& \multicolumn{3}{c} {$r=20$}&&\multicolumn{3}{c}   {$r=15$}\\
       % \cmidrule(r){3-6}           \cmidrule(r){8-11}  \\
                \cline{4-6}        \cline{7-10}
 & \multirow{8}{*}{$50$}
     &   H-PWLS   & 82&2.0 &0.6   &&77 & 2.7 &0.7   \\
    & & PWLS &31 &11.1 & 1.7        && 23 & 14.7 &1.8 \\
     &  & HIPOD &38 &9.3 & 3.2        & & 33 & 10.2 &5.2  \\  \cline{4-10}
    & && \multicolumn{3}{c} {$r=10$}&&\multicolumn{3}{c}   {$r=5$}\\
                \cline{4-6}        \cline{7-10}
      &  & H-PWLS&74 & 3.0 &0.8     &&12 &18.6 &1.6  \\
     &   & PWLS&13 & 19.2 &2.3     &&2 &37.6 &4.7 \\
     &   & HIPOD&22 & 14.1 &5.8    &&2 &35.3 &5.6 \\
 \cline{1-10}
  \multirow{16}{*}{\rotatebox[origin=c]{90}{Case 2}}
       &&& \multicolumn{3}{c} {$r=20$}&&\multicolumn{3}{c}   {$r=15$}\\
               \cline{4-6}        \cline{7-10}
&  \multirow{8}{*}{$15$}
        & H-PWLS   &88 &  3.5 & 0.5     &&91 &1.1& 0.6 \\
       & & PWLS     &90  & 2.5 &0.6    &&68 &3.8 &0.4 \\
      &  & HIPOD    &94 &0.7 &2.2     &&84 &1.8 &2.2 \\  \cline{4-10}
   & && \multicolumn{3}{c} {$r=10$}&&\multicolumn{3}{c}   {$r=5$}\\
       \cline{4-6}        \cline{7-10}
         && H-PWLS & 75 &2.9 & 0.7       &&14 &17.3 &1.5  \\
         && PWLS  & 41  &8.2  &0.7       && 7 &25.0 &2.1 \\
         && HIPOD & 63  &4.8 & 2.2       && 7 &23.7 &2.2 \\ \cline{3-10}
   &&& \multicolumn{3}{c} {$r=20$}&&\multicolumn{3}{c}   {$r=15$}\\
       % \cmidrule(r){3-6}           \cmidrule(r){8-11}  \\
       \cline{4-6}        \cline{7-10}
  &  \multirow{8}{*}{$50$}
       & H-PWLS & 82  &2.0 & 0.2   &&80& 2.2 &0.2   \\
       &&PWLS    &70  &3.4  &0.4          &&64& 4.4 &0.5 \\
      & &HIPOD   &82  &2.0  &2.2       &&72& 3.2 &2.2  \\  \cline{4-10}
    &&& \multicolumn{3}{c} {$r=10$}&&\multicolumn{3}{c}   {$r=5$}\\
           \cline{4-6}        \cline{7-10}
       && H-PWLS&67& 4.2& 0.3     &&13& 20.9& 1.1  \\
       && PWLS  &43& 7.8 &0.7     &&8 &28.3& 2.1 \\
       && HIPOD &59 &5.3& 2.2    &&8 &28.0& 2.1 \\     \hline\hline
\end{tabular}
\end{center}
\end{table}

From Table \ref{table-outlier-hetero}, we see that the H-PWLS significantly outperforms both
 the PWLS and HIPOD in all settings; it has much higher outlier detection rate measured in JD, and at
  the same time has much smaller masking error and swamping error measured in M and S.
   Especially, when $r=20$,
the H-PWLS detects almost  all the outliers correctly with very small swamping error,
whereas neither the PWLS nor the HIPOD works well. The H-PWLS is also illustrated in
 detail in Figure \ref{Fig:hetero} using the data of the first repetition from the setting where $r=10$ and $p=15$.
 In addition, H-PWLS still performs consistently better than both the PWLS and HIPOD
 when the variance function is mis-specified. It means that the H-PWLS is relatively robust
 to the mis-specification of the variance function.

\subsection{Real data applications} \label{sec-realdata}

We now apply the PWLS  to three real datasets:
Coleman data, Salinity data, and Real Estate data.
We use the random weighting method to select the tuning parameter
and produce outlier probabilities for all observations.

\subsubsection{Coleman Data}

The Coleman data were obtained from a sample studied by \cite{coleman:1966} and
further analyzed by \cite{MostellerTukey:1977} and \cite{rousseeuw.leroy:1987}.
The data include six different measurements of 20 Schools from the Mid-Atlantic
and New England States. These measurements are: (1) staff salaries per pupil ({\it salaryP}),
(2) percent of white-collar fathers ({\it fatherWc}), (3) socioeconomic status composite deviation ({\it sstatus}),
(4) mean teacher's verbal test score ({\it teacherSc}),
(5) mean mother's educational level ({\it motherLev}), and (6) verbal mean test score of all six graders (the outcome variable $y$).
One wants to estimate the verbal mean test score from 5 other measurements using linear regression.

\begin{figure}[thp]
  \caption{First column: Coleman Data; Second column: Salinity Data;  Third column: Real Estate Data.
   Top row: PLWS weight solution paths; Bottom row: Outlier probability paths}\label{fig-data}
\centering
$$
 \scalebox{0.38} [0.33]{\includegraphics{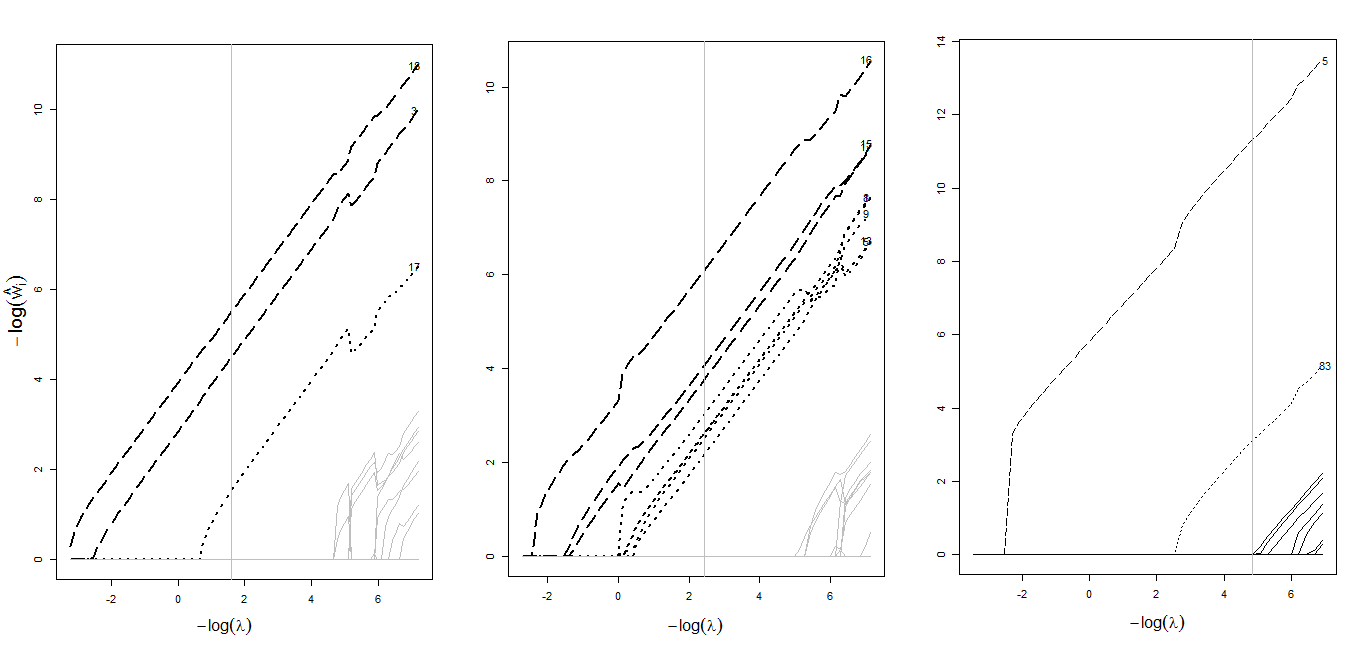}}
 $$
 $$
 \scalebox{0.38} [0.33]{\includegraphics{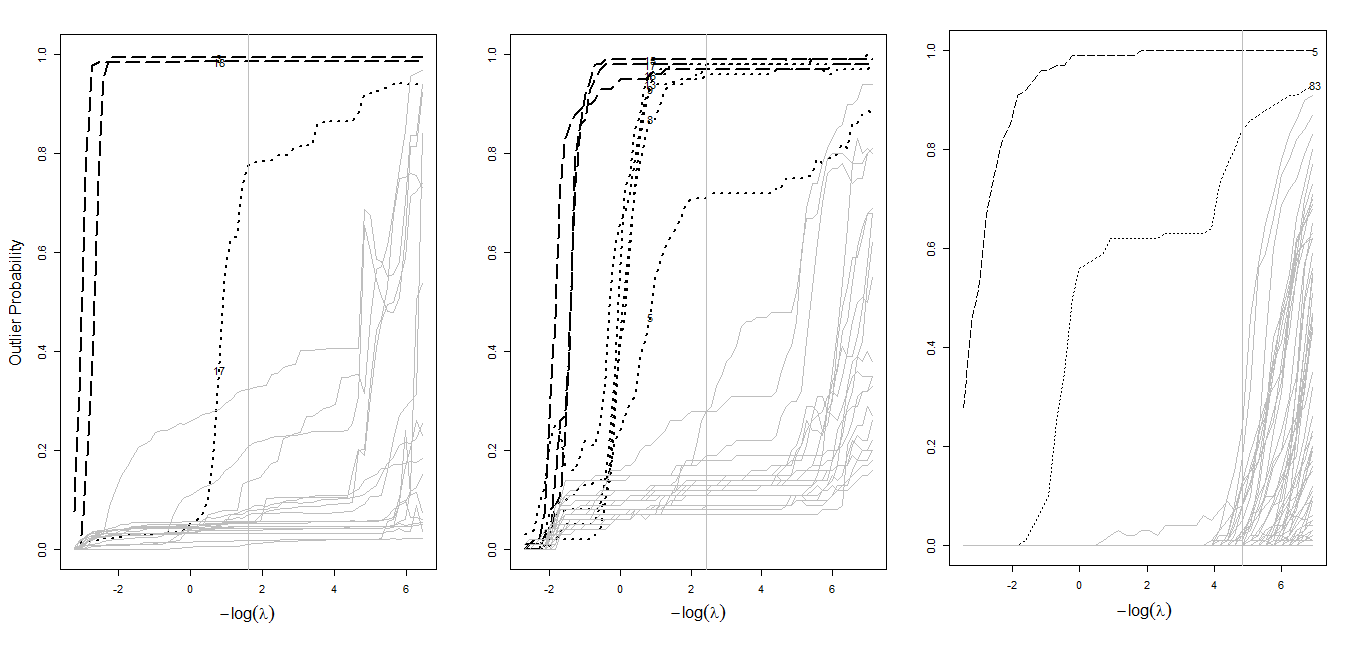}}
 $$
\end{figure}

The PWLS analysis results are plotted in Figure \ref{fig-data},
where both the weight solution path
and outlier probabilities along a sequences of $\lm$ are plotted in
the top and bottom panels, respectively.
 The vertical line is at the selected tuning parameter
$\widehat{\lambda}$.  This also applies for all plots Figure \ref{fig-data}.
 The PWLS weights solution path (top-left panel) tells how weight
$\widehat{w}_i(\lambda)$  changes with tuning parameter $\lambda$ for observation $i$, $1\leq i \leq 20$.
From the PWLS analysis, we suggest to downweight observations 3rd, 17th, and 18th
in the regression analysis.

The outlier probability plot (bottom-left panel) shows
 the  trajectory  of outlier probability $\widehat{P}^o_i(\lambda)$ for those $\lambda$.
 Both the 3rd and 18th observations are very likely to be outliers since the outlier probabilities
are $0.99$ around the vertical line. Comparing with other observations with less than 50\% outlier probabilities,
 the 17th observation stands out with a much higher outlier probability of $0.86$ around the vertical line.
 However,  the HIPOD claims four additional outliers.

The  corresponding regression
estimation results are summarized at the top of  Table \ref{tab-data} and it shows that
both {\it sstatus} and  {\it teacherSc} are positively associated with the outcome,
while {\it motherLev} is negatively associated with it.
Results from both the HIPOD is also listed as a comparison.
The HIPOD turns to choose more outliers than the PWLS.

\begin{table}[htp]
\begin{center}
 \caption{Robust Estimation Results by PWLS and HIPOD  for all three data sets}\label{tab-data}
\begin{tabular}{llrrrrrr}\\ \hline \hline
%\multirow{4}{*}{\rotatebox[origin=c]{90}{Est.}} &
\multirow{4}{*}{{\it Coleman}} &
      & Int.  & {\it salaryP}   &{\it fatherWc}  &{\it sstatus}   &{\it teacherSc} & {\it motherLev} \\
            \cline{3-8}
       &PWLS     & 32.097 & $-1.644$&   0.079&  0.656 &   1.110& $-4.149$ \\
        &HIPOD    & 1.928 & $-1.613$ &   0.030&   0.609 &    1.485 &   $-0.489$ \\
     %   &FS     & 29.750&   $-1.203$&     0.082&    0.659&      1.098 &    $-3.898$ \\
        \hline
        \multirow{4}{*}{{\it Salinity}} &
                & Int.  & $x_1$ & $x_2$  & $x_3$   &  &\\
           \cline{3-8}
  &PWLS         &  16.913 & 0.711& $-0.134$&  $-0.571$ &  &\\
  &HIPOD      & 0.004 & 1.164 & $-0.076$ & $-0.049$ &   &\\
 % &FS        & 36.650& 0.389& $-0.114$& $-1.307$ & &\\
  \hline
  \multirow{4}{*}{{\it Real Estate}} &
       & Int.  &Year  &Area & Story & Land & \\
            \cline{3-8}
       &PWLS &  1.228&       0.006 &  0.377 & 0.042  &    0.061  &\\
         &HIPOD &  1.002  &    0.010& 0.296 &0.008    & 0.082 &\\
    %      &FS &  1.227   &     0.006 &  0.377 &  0.042   &    0.061 &\\
\hline\hline
\end{tabular}
\end{center}
\end{table}

\subsubsection{Salinity Data}

The Salinity data consists of 28 sample points of water salinity (i.e., its salt concentration) and river
discharge taken in North Carolina's Pamlico Sound. The data have four measurements:
(1) lagged salinity ($x_1$), (2) trend ($x_2$), (3) discharge ($x_3$), and the outcome variable,
salinity ($y$). The data were analyzed by \cite{ruppert.carroll:1980} and \cite{carroll.ruppert.1995}.
\cite{carroll.ruppert.1995} found that the 5th observation was masked by the 13th and 16th observations,
which are corresponding to two periods of very heavy discharge.

The PWLS analysis results  are reported in second column of  Figure \ref{fig-data}.
The weight solution path (top-middle panel) suggests eight observations should be
downweighted in the the regression analysis, the 1st, 5th, 8th, 9th,
  13th, 15th, 16th, and 17th. Since there are 8 suspected outliers, some subgroup analysis on them might also be helpful.
To understand these 8 suspected outliers in more detail, we examined the outlier probability
plot (bottom-middle panel)  carefully. All, except observation 5,  among  these 8 suspected outliers have
 very high outlier probabilities ($>0.9$).  The outlier probability of the 5th observation is about 0.7,
 is also much higher than and one for the remaining 20 observations ($< 0.3$).

Using the HIPOD, 15 out of 28 samples (observations
 4,  6,  7,  9, 10, 11, 13, 14, 15, 17, 18, 19, 21, 23 and 24)  are outliers,
 while observation 5 is masked.

%Thus, both FS and HIPOD generates considerably more outliers than the PWLS.

The  PWLS regression
estimation results and its corresponding comparisons
with HIPOD are summarized in the middle of Table \ref{tab-data}. It shows that $x_1$ (lagged salinity) is positively
 associated with the outcome, while $x_3$ (discharge) is negatively associated with it.

%Again, the result from PWLS shares more similarity with one from FS than one from the IPOD.

\subsubsection{Real Estate Data}
The Real Estate data is taken from a Wake County, North Carolina real estate database in 2008 \citep{woodard.leone:2008}.
%Wake County boasts a 31.18\% growth in population since 2000, with a population of approximately 823,345 residents.
This original data set includes
11 recorded variables for 100 randomly selected residential properties in the Wake County registry
denoted by their real estate ID number.

The aim of the study is to predict the total assessed value of the property (logTotal: log(total value/10,000))
using  four variables:
the listed year in which the structure was built (Year: year built-1900),
  the area of the floor plan in square feet (Area: in 1000 square feet),
  the number of stories the structure has (Stories),
and  the assessed value of the land (Land, in \$10,000). Three properties (ID number 36, 50 and 58)
are removed from the study since they have $0$ land values.

The PWLS analysis results are reported in the third column of Figure \ref{fig-data}.
From the weight solution path (top-right panel), observations 5 and 83 (real estate ID number 86) are claimed to be outliers.
Both of those two observations  have considerably larger outlier probabilities ( 1.0 and 0.84)
 than the other observations (less than 0.2). See the bottom-right panel at Figure  \ref{fig-data}.
  The HIPOD  obtains the same outlier set as the PWLS.

Robust estimation results from PWLS and HIPOD
are  summarized
at the bottom of  Table \ref{tab-data}  and it shows  both methods
providing consistent results for this data set.

\section{Discussion}

In this manuscript, we propose a new approach to analyzing data with contamination. By assigning each observation an individual weight
and incorporating an adaptive lasso-type penalty on the log-transformation of the weight vector, the aPWLS is able to perform outlier detection and robust regression.

However, like any existing penalized approach, the problem of tuning parameter selection in the aPWLS is notorious. On the one hand, the selection of tuning parameter plays an extremely important role, because it determines the number of suspected outliers. On the other hand, there is no gold standard on how to select the tuning parameter. In this manuscript, we propose two tuning methods, BIC and random weighting.
 The BIC was used widely in the literature and also used by \cite{she.owen:2011} for tuning IPOD. Random weighting is a new idea for tuning parameter selection. Based on our limited numerical experience, there is not much difference in the performance between these two tuning methods, but the random weighting method can provide for each observations the probability of its being an outlier. As demonstrated using the HBK data and two real datasets, these outlier probabilities are useful for visualizing the performance of the PWLS as the tuning parameter changes.

Robust regression with variable selection has attracted much attention lately in
high-dimensional data analysis.
See, for example, the adaptive Lasso penalty under $\ell_1$ loss in \cite{wang.li.jiang:2007},
 Huber's loss in \cite{lambert.zwald:2011}
 and  the least trimmed squares loss
in \cite{alfons.croux.gelper:2013}. A huge literature review on variable selection
can  be found in \citealp{hastie.tibshirani.ea:2009}.
Actually, we could also conduct variable selection and outlier detection simultaneously,
by adding an extra penalty on the regression coefficients, say $\lambda_2\sum_{j=1}^p|\beta_j|$,
to the objective function of (\ref{eq:pwls}).

Moreover, it is important to point out that the extended aPWLS proposed in Section 4 is actually a
variation of the classical WLS aiming for fitting heterogeneity data, with the main
goal being outlier detection. We consider the variance function linear models (VFLM)
in Section 4, which is more general than the heterogenous model behind the classical WLS,
and we can further extend the aPWLS to any variance function models.

\section*{Appendix}

\noindent{\bf Proof of Theorem \ref{thm:m-est}}

  The proof is similar to She and Owen (2011).
  Due to the  convexity properties,
  we only need to check the equivalence of joint KKT functions under
   both \eqref{eq:rho-con} and \eqref{eq:pwls-con}.

 We first consider joint KKT equations under \eqref{eq:rho-con},
  $$
  \left\{
  \begin{array}{l}
  \sum_{i=1}^n\bx_i'\psi ((y_i-\bx_i'\bbeta)/\sigma; \lm )=0,\\
  \dfrac{\partial}{\partial\sigma} \left(\sum_{i=1}^n\rho ( (y_i-\bx_i'\bbeta)/\sigma; \lm)
        +2cn\log\sigma \right )=0.
  \end{array}
  \right.
  $$
%  We can check
% $$\dfrac{\partial}{\partial\sigma}(\rho(t/\sigma, \lm)
% = \left\{ \begin{array}{ll}
%  \dfrac{\partial}{\partial\sigma}(\lm\log(|t|/\sigma)+\lm/2
%  & {\rm if~}|t|>\sqrt{\lm}\sigma\\
%    \dfrac{\partial}{\partial\sigma}(t^2/2\sigma^2)
%  & {\rm if~}|t|\le\sqrt{\lm}\sigma\\
%    \end{array}
%  \right.$$
We can check
 \bel{appendix-eq-1}
 \dfrac{\partial}{\partial\sigma}( \rho(t/\sigma, \lm)+cn\log(\sigma) )
 = \left\{ \begin{array}{ll}
  -\lm/\sigma+2 cn/\sigma
  & {\rm if~}|t|>\sqrt{\lm/2}\sigma,\\
   -2t^2/\sigma^3+2 cn/\sigma
  & {\rm if~}|t|\le\sqrt{\lm/2}\sigma.\\
    \end{array}
  \right.
 \eel
 Replace $t$ by $y_i-\bx_i'\bbeta$ and set \eqref{appendix-eq-1} to be $0$,
 we have
 $$cn-\sum_{i\in \widehat{\mathcal{O}}} \lm/2-\sum_{i\in \widehat {\mathcal{O}}^c} (y_i-\bx_i'\hbbeta)^2/\sigma^2=0,$$
 where $\widehat{\mathcal{O}}=\{1\le i\le n: |y_i-\bx_i'\hbbeta|>\sqrt{\lm/2}\sigma\}$ and
 $\widehat {\mathcal{O}}^c=\{1\le i\le n: |y_i-\bx_i'\hbbeta|\le \sqrt{\lm/2}\sigma\}$.
  Denote $r_i=y_i-\bx_i'\bbeta$ and  $\br=(r_1,\cdots, r_n)'$.
  Then $\widehat\br_{\widehat {\mathcal{O}}^c}$ is the sub-vector of $\widehat \br$
  for all observations in $\mathcal{O}^c$. Thus, we have
\bel{eq-proof-sig}
\widehat \sigma^2=\|\widehat \br_{\widehat {\mathcal{O}}^c}\|_2^2/(cn-(\lm/2)\sharp\{\widehat{\mathcal{O}}\}),
 \eel
 where $ \sharp\{\widehat{\mathcal{O}}\}$ is the cardinal value of set $\widehat{\mathcal{O}}.$
 We now consider joint KKT of the penalized objective function in  \eqref{eq:pwls-con}.
From the derivative on $\bw$,
  $$
2(w_i/\sigma^{2})(y_i-\bx_i'\bbeta)^2+\lm\sgn(\log(w_i))(1/w_i)=0.
$$
We obtain
\bel{eq-proof-w}
w_i= \left\{ \begin{array}{ll}
 \sqrt{\lm/2}(\sigma/|\widehat r_i|)
  & {\rm if~}|\widehat r_i|>\sigma\sqrt{\lm/2},\\
   1
  & {\rm if~}|\widehat r_i|\le \sigma\sqrt{\lm/2}.\\
    \end{array}
  \right.
\eel
From the derivative on $\sigma$,
\bel{eq-proof-sig2}
cn\sigma^2=\sum_{i=1}^n w_i^2 (y_i-\bx_i\hbbeta)^2=\sum_{i\in\widehat {\mathcal{O}}^c}\widehat r_i^2+
  \sum_{i\in \widehat{\mathcal{O}}}(\widehat w_i^2 \widehat r_i^2).
\eel
Combining with \eqref{eq-proof-w} and \eqref{eq-proof-sig2}, we can
also obtain \eqref{eq-proof-sig}.

Finally,
plugging in \eqref{eq-proof-w} in  \eqref{eq:pwls-con},
we are able to obtain the concomitant M-estimation $\rho$ function
in  \eqref{eq:rho-con}. $\Box$

\bibliographystyle{agsm}
\bibliography{allrefers-Gao17AUG2015}
%\bibliography{allrefers-Gao1Nov2015}

\end{document}